\documentclass[twocolumn,english]{svjour3}
\usepackage[T1]{fontenc}
\usepackage[latin9]{inputenc}
\usepackage{url}
\usepackage{amsmath}
\usepackage{amssymb}
\usepackage{esint}

\makeatletter
\RequirePackage{fix-cm}

\smartqed  

\makeatother

\usepackage{babel}
\begin{document}
\title{Annular structures in perturbed low mass disc-shaped gaseous nebulae
I : general and standard models}
\titlerunning{Annular structures in gaseous nebulae I : general and standard models}
\author{Vladimir Pletser }
\authorrunning{V. Pletser}
\institute{lnstitut d'Astronomie et de Geophysique G.Lemaitre, Catholic University
of Louvain, Louvain-la-Neuve, Belgium \\
\emph{Present address:} Blue Abyss, Newquay, Cornwall, United Kingdom}

\maketitle
Email: pletservladimir@gmail.com
\institute{ORCID 0000-0003-4884-3827}
\date{Received: date / Accepted: date}
\begin{abstract}
This is the first of two papers where we study analytical solutions
of a bidimensional low mass gaseous disc slowly rotating around a
central mass and submitted to small radial periodic perturbations.
Hydrodynamics equations are solved for the equilibrium and perturbed
configurations. A wave-like equation for the gas perturbed specific
mass is deduced and solved analytically for several cases of exponents
of the power law distributions of the unperturbed specific mass and
sound speed. It is found that, first, the gas perturbed specific mass
displays exponentially spaced maxima, corresponding to zeros of the
radial perturbed velocity; second, the distance ratio of successive
maxima of the perturbed specific mass is a constant depending on disc
characteristics and, following the model, also on the perturbation's
frequency; and, third, inward and outward gas flows are induced from
zones of minima toward zones of maxima of perturbed specific mass,
leading eventually to the possible formation of gaseous annular structures
in the disc. The results presented may be applied in various astrophysical
contexts to slowly rotating thin gaseous discs of negligible relative
mass, submitted to small radial periodic perturbations. 

\end{abstract}

\section{Introduction\label{sec:Introduction}}

Discs play an important role in astrophysics (see e.g., \cite{Latter et al 2018}).
Protostellar discs are found around certain categories of young stars.
Dynamical accretion discs intervene in the feeding process of massive
stars by less massive ones in some binary systems. Galaxies have often
the shape of a disc surrounding a central bulge. Planetary rings form
discs around giant planets. Furthermore, it is generally believed
that the planetary system and the regular satellites systems originate
from disc-shaped nebulae surrounding the proto-Sun and the giant proto-planets.
The disc stage is thus an important step in some systems evolution.
Depending on disc and central mass characteristic and on their mutual
relative importance, different kind of structures may appear in discs:
bars, spiral arms, rings. Theory of disc dynamical evolution in an
astrophysical context may be applied to other situations, for example
the theory of spiral density waves of galactic arms was successfully
applied to models of planetary rings (see e.g., \cite{Shu1984}).
Characteristics of a disc that may influence its evolution are self-gravity,
thermal pressure, interaction with magnetic fields, rotation and viscosity.
In this paper, we show that annular rings may appear under certain
circumstances in slowly rotating thin low mass gaseous discs, where
self-gravitation, viscous, magneto-hydrodynamics effects and azimuthal
perturbations can be neglected. We study the behaviour of a nebular
disc taken away from equilibrium by small radial periodic perturbations,
extending the classical Jeans' model of an uncompressible isothermal
stationary nebula submitted to perturbations. Although initially intended
for protoplanetary discs, the results of the investigations presented
here can be applied to any thin gaseous disc that can be described
by the model considered. Hypotheses on the model are discussed in
section 2. We consider in section 3 a differentially rotating thin
gaseous axisymmetric nebula undergoing polytropic transformations
of index $\gamma$ and departing from equilibrium because of small
radial periodic perturbations. Physical characteristics of the nebular
disc are supposed continuous and have power law dependencies on the
radial distance $r$, in particular for the specific mass $\rho\sim r^{d}$
and sound speed $c\sim r^{\frac{s}{2}}$. Equations describing the
hydrodynamic model are solved for the equilibrium and perturbed configurations,
where the perturbations are assumed small enough for the equations
to be linearized. A wave-like equation is deduced for the nebula perturbed
gas specific mass and expressions of the gas radial velocity and specific
mass flux momentum are found in function of the gas specific mass.
Looking for solutions yielding annular gaseous structures to appear
in the disc, these equations are solved analytically in Section 4
for two particular models ($d=0$ and $d<2\left(2\gamma-1\right)$;
$s=2$) and for a third general case ($d=\left(s-2\right)$; $s<2$)
for small frequencies. A particular case of the latter with $s=-1$
and $d=-3$, called the \textquotedbl standard model\textquotedbl ,
is briefly introduced. Expressions of the distance ratios $\beta$
of the maxima of the perturbed gas specific mass are also deduced.
Profiles of the perturbed specific mass and velocity are presented
in Section 5 and the possible formation of annular structures are
discussed. We are not aware of previous similar general analytical
resolutions, although particular cases were treated in \cite{Nowotny1979}.
In a second paper, we explore analytical solutions for two other general
models, including a polytropic case. Both papers are reworked excerpts
of \cite{Pletser1990}.

\section{Hypotheses on the disc model\label{sec:2 Hypotheses on the disc model}}

The mass of a primeval nebula is a key factor in deciding on its later
evolution: either the nebula mass is large, typically greater than
or close to the central mass, and sub-regions of the nebula of large
specific mass may undergo local collapse, or the primeval nebula mass
is low, typically a few percents of the central mass and gravitational
instabilities may never develop in this case \cite{Lin1986}. In some
theories of protoplanetary nebula formation (see e.g., \cite{Black_=000026_Matthews1985}),
viscous friction plays an important role in inducing an inward flow
of accretion material onto the central primary and causing the conversion
of kinetic into thermal energy to be the dominant heat source \cite{Morfill_etal1985}.
However, the epoch at which the viscous friction becomes the predominant
effect is critical in a nebula history. After an initial collapse
phase, a low mass rotating nebula can achieve a stationary equilibrium
without considering turbulent friction \cite{Bodenheimer_=000026_Tscharnuter1979}.
On the other hand, an axisymmetric equilibrium configuration was shown
to be unstable against non-axisymmetric perturbations, the result
being a binary system \cite{Rozyczka_etal1980a,Rozyczka_etal1980b}.
Furthermore, friction processes are not always able to produce a central
object surrounded by a disc-shaped nebula \cite{Morfill_etal1985}.
Therefore, it is reasonable to assume, within the low mass nebula
hypothesis, that there was a period in a nebula history during which
the viscous friction may not have been the predominant process governing
the disc evolution, independently of further evolution where the viscous
effects may have become predominant. The problem of transfer of angular
momentum from the central mass to outer parts of the disc is not addressed
here, as it depends on viscous processes (see e.g., \cite{Cassen_etal1985}). 

We consider the model of a disc-shaped gaseous nebular disc of mass
$M_{d}$ in a slow rotation around a central mass $M^{*}$, supposed
spherical. The disc mass is negligible in front of the central mass
$M_{d}<<M^{*}$ and the disc thickness is small compared to its radius.
The nebula is assumed to be composed of gas only, the presence of
nebular dust being neglected. Self-gravitation, magneto-hydrodynamics
and viscous effects in the disc are not considered (although, the
viscous force is included in the general equations of section 3, but
neglected further in section 4). 

The effects of small periodic radial perturbations on the disc are
studied, without any coupling to non-radial perturbations. This last
hypothesis is somehow controversial as there is a large body of work
(see e.g, \cite{Goldreich =000026 Tremaine 1979,Blumenthal et al 1984,Lin et al 1990,Lubow =000026  Pringle 1993}
and references therein) that consider coupling between radial and
azimuthal perturbations, typically through the Coriolis force. However,
for slowly rotating discs, i.e., for which the angular frequency of
rotation $\Omega$ is smaller than the perturbation periodic angular
frequency $\omega$, with $\Omega<<\omega$, the error committed by
ignoring the azimuthal perturbations would be small. Although this
approximation is strictly speaking incorrect, as we will be interested
further in the radial distributions of the perturbed variables, the
small azimuthal effects are ignored in a first approach in this study.
Nevertheless, the nebula model equations are deduced for the radial,
azimuthal and vertical components, and we show that azimuthal perturbations
are negligible for a slowly rotating disc and vertical perturbations
are non-existent for an inviscid disc.

\section{Disc hydrodynamic model \label{sec:3 Disc-hydrodynamic-model}}

The motion of the gas of specific mass $\rho$ is described in spatial
Eulerian coordinates by the vectorial Navier-Stokes equation, which
relates for an unit volume of gas, the inertia force (sum of the time
derivative, denoted by an upper dot, of the vectorial velocity $\boldsymbol{v}$
and the advection term), the gradients of the pressure $p$ and the
gravitational potential $V$, and the viscous forces
\begin{equation}
\rho\dot{\boldsymbol{v}}+\rho\left(\boldsymbol{v}\cdot\nabla\right)\boldsymbol{v}+\nabla p+\rho\nabla V=\rho\nu\boldsymbol{\Delta}\boldsymbol{v}\label{eq:1}
\end{equation}
where $\nu$ is the kinematic viscosity, $\nabla$ and $\boldsymbol{\Delta}$
the gradient and vectorial Laplacian operators. This equation is complemented
by the continuity and Poisson equations
\begin{equation}
\dot{\rho}+\nabla\cdot\left(\rho\boldsymbol{v}\right)=0\label{eq:2}
\end{equation}
\begin{equation}
\Delta V=4\pi G\rho\label{eq:3}
\end{equation}
where $\nabla\cdot$ and $\Delta$ are the divergence and scalar Laplacian
operators and $G$ the gravitational constant. The disc gravitational
potential is neglected in front of the central mass gravitational
potential and the viscosity $\nu$ is assumed constant in the disc. 

At dynamic equilibrium, the stationary model is described by 
\begin{equation}
\rho_{o}\left(\boldsymbol{v_{0}}\cdot\nabla\right)\boldsymbol{v_{0}}+\nabla p_{0}+\rho_{0}\nabla V_{0}=\rho_{0}\nu\boldsymbol{\Delta}\boldsymbol{v_{0}}\label{eq:4}
\end{equation}
\begin{equation}
\nabla\cdot\left(\rho_{0}\boldsymbol{v}_{0}\right)=0\label{eq:5}
\end{equation}
\begin{equation}
\Delta V_{0}=4\pi G\rho_{0}\label{eq:6}
\end{equation}
where the index $0$ denotes the equilibrium characteristics. Allowing
for small radial periodic perturbations to take the model away from
equilibrium, the linearized perturbed equations read, after simplification
by the equilibrium equations \eqref{eq:4} to \eqref{eq:6},
\begin{align}
 & \rho_{0}\dot{\boldsymbol{v_{1}}}+\rho_{0}\left(\left(\boldsymbol{v_{1}}\cdot\nabla\right)\boldsymbol{v_{0}}+\left(\boldsymbol{v_{0}}\cdot\nabla\right)\boldsymbol{v_{1}}\right)+\rho_{1}\left(\boldsymbol{v_{0}}\cdot\nabla\right)\boldsymbol{v_{0}}\nonumber \\
 & +\nabla p_{1}+\rho_{1}\nabla V_{0}=\rho_{0}\nu\boldsymbol{\Delta}\boldsymbol{v_{1}}+\rho_{1}\nu\boldsymbol{\Delta}\boldsymbol{v_{0}}\label{eq:7}
\end{align}

\begin{equation}
\dot{\rho_{1}}+\nabla\cdot\left(\rho_{1}\boldsymbol{v_{0}}\right)+\nabla\cdot\left(\rho_{0}\boldsymbol{v_{1}}\right)=0\label{eq:8}
\end{equation}
\begin{equation}
\Delta V_{1}=4\pi G\rho_{1}\label{eq:9}
\end{equation}
where indexes $1$ denote the perturbed characteristics. As the model
is plane and axisymmetric, these equations are solved in a cylindrical
polar reference frame. Considering that the equilibrium characteristics
depend only on the radial distance $r$ and that the perturbed characteristics
depend on $r$ and on the time $t$, the equilibrium and perturbed
gas vectorial velocities are written respectively
\begin{align*}
\boldsymbol{v_{0}} & =\left(0,v_{0}\left(r\right),0\right)\\
\boldsymbol{v_{1}} & =\left(v_{1}\left(r,t\right),u_{1}\left(r,t\right),w_{1}\left(r,t\right)\right)
\end{align*}
At dynamical equilibrium, the radial and azimuthal components of the
Navier-Stokes equation \eqref{eq:4} and the Poisson equation \eqref{eq:6}
read, with the prime sign $^{\prime}$ denoting $\partial\,/\partial r$,
\begin{equation}
\frac{\rho_{o}v_{0}^{2}}{r}-p_{0}^{\prime}-\rho_{0}V_{0}^{\prime}=0\label{eq:10}
\end{equation}

\begin{equation}
\rho_{0}\nu\left(v_{0}^{\prime\prime}+\frac{v_{0}^{\prime}}{r}-\frac{v_{0}}{r^{2}}\right)=0\label{eq:11}
\end{equation}

\begin{equation}
V_{0}^{\prime\prime}+\frac{V_{0}^{\prime}}{r}=4\pi G\rho_{0}\label{eq:12}
\end{equation}
The Navier-Stokes equations \eqref{eq:7} for the perturbed radial
component reads
\begin{align}
 & \rho_{0}\dot{v_{1}}-\frac{v_{0}}{r}\left(\rho_{1}v_{0}+2\rho_{0}u_{1}\right)+p_{1}^{\prime}+\rho_{0}V_{1}^{\prime}+\rho_{1}V_{0}^{\prime}\nonumber \\
 & =\rho_{0}\nu\left(v_{1}^{\prime\prime}+\frac{v_{1}^{\prime}}{r}-\frac{v_{1}}{r^{2}}\right)\label{eq:13}
\end{align}
The second and third terms of \eqref{eq:13} can be simplified as$\rho_{1}v_{0}>>2\rho_{0}u_{1}$
(see Appendix A), yielding 
\begin{equation}
\rho_{0}\dot{v_{1}}-\rho_{1}\frac{v_{0}^{2}}{r}+p_{1}^{\prime}+\rho_{0}V_{1}^{\prime}+\rho_{1}V_{0}^{\prime}=\rho_{0}\nu\left(v_{1}^{\prime\prime}+\frac{v_{1}^{\prime}}{r}-\frac{v_{1}}{r^{2}}\right)\label{eq:13-1}
\end{equation}
The Navier-Stokes equations \eqref{eq:7} for the perturbed azimuthal
and vertical components become

\begin{align}
 & \rho_{0}\dot{u_{1}}+\rho_{0}v_{1}\left(v_{0}^{\prime}+\frac{v_{0}}{r}\right)=\rho_{1}\nu\left(v_{0}^{\prime\prime}+\frac{v_{0}^{\prime}}{r}-\frac{v_{0}}{r^{2}}\right)\nonumber \\
 & +\rho_{0}\nu\left(u_{1}^{\prime\prime}+\frac{u_{1}^{\prime}}{r}-\frac{u_{1}}{r^{2}}\right)\label{eq:14}
\end{align}

\begin{equation}
\rho_{0}\dot{w_{1}}=\rho_{0}\nu\left(w_{1}^{\prime\prime}+\frac{w_{1}^{\prime}}{r}\right)\label{eq:14-1}
\end{equation}
The continuity and Poisson equations \eqref{eq:8} and \eqref{eq:9}
read
\begin{equation}
\dot{\rho_{1}}+\frac{\rho_{0}v_{1}}{r}+\rho_{0}^{\prime}v_{1}+\rho_{0}v_{1}^{\prime}=0\label{eq: 15}
\end{equation}
\begin{equation}
V_{1}^{\prime\prime}+\frac{V_{1}^{\prime}}{r}=4\pi G\rho_{1}\label{eq:16}
\end{equation}
This set of equations is completed by a gas state equation. The nebula
gas is approximated by a perfect gas undergoing polytropic transformations
of index $\gamma$, assumed to be constant throughout the disc. Denoting
the local sound speed by $c$, the pressure at equilibrium reads
\begin{equation}
p_{0}=\frac{c_{0}^{2}\rho_{0}}{\gamma}\label{eq:17}
\end{equation}
Using the gas polytropic relation, $p\,/\rho^{\gamma}=constant$,
the linearized perturbed pressure reads
\begin{equation}
p_{1}=\frac{c_{0}^{2}\rho_{1}}{\gamma}+2\frac{c_{0}c_{1}\rho_{0}}{\gamma}=c_{0}^{2}\rho_{1}\label{eq:18}
\end{equation}
Expressions of the gas circular velocity at equilibrium are found
from the radial and azimuthal components of the Navier-Stokes equation
\eqref{eq:10} and \eqref{eq:11} and are given in Appendix B. 

Solving for the gas perturbed specific mass $\rho_{1}$ and perturbed
radial velocity $v_{1}$, the equation \eqref{eq:13-1}, with \eqref{eq:10},
\eqref{eq:17} and \eqref{eq:18}, reads
\begin{align}
 & \dot{v_{1}}+\frac{c_{0}^{2}}{\rho_{0}}\left(\rho_{1}^{\prime}+\rho_{1}\left(\left(\frac{\gamma-1}{\gamma}\right)\frac{\left(c_{0}^{2}\right)^{\prime}}{c_{0}^{2}}-\frac{\rho_{0}^{\prime}}{\gamma\rho_{0}}\right)\right)+V_{1}^{\prime}\nonumber \\
 & =\nu\left(v_{1}^{\prime\prime}+\frac{v_{1}^{\prime}}{r}-\frac{v_{1}}{r^{2}}\right)\label{eq:19}
\end{align}
Taking the time derivative of \eqref{eq: 15} and introducing \eqref{eq:16}
and \eqref{eq:19} yield
\begin{align}
 & \ddot{\rho_{1}}-c_{0}^{2}\left(\rho_{1}^{\prime\prime}+\rho_{1}^{\prime}\left(\left(\frac{2\gamma-1}{\gamma}\right)\frac{\left(c_{0}^{2}\right)^{\prime}}{c_{0}^{2}}-\frac{\rho_{0}^{\prime}}{\gamma\rho_{0}}+\frac{1}{r}\right)\right.\nonumber \\
 & \left.+\rho_{1}\left(\left(\frac{\gamma-1}{\gamma}\right)\frac{\left(c_{0}^{2}\right)^{\prime\prime}}{c_{0}^{2}}+\frac{\left(c_{0}^{2}\right)^{\prime}}{c_{0}^{2}}\left(\left(\frac{\gamma-1}{\gamma}\right)\frac{1}{r}-\frac{\rho_{0}^{\prime}}{\gamma\rho_{0}}\right)\right.\right.\nonumber \\
 & \left.\left.-\frac{1}{\gamma\rho_{0}}\left(\rho_{0}^{\prime\prime}+\rho_{0}^{\prime}\left(\frac{1}{r}-\frac{\rho_{0}^{\prime}}{\rho_{0}}\right)\right)+\frac{4\pi G\rho_{0}}{c_{0}^{2}}\right)\right)\nonumber \\
 & =\rho_{0}^{\prime}V_{1}^{\prime}-\frac{1}{r}\frac{\partial}{\partial r}\left(r\rho_{0}\nu\left(v_{1}^{\prime\prime}+\frac{v_{1}^{\prime}}{r}-\frac{v_{1}}{r^{2}}\right)\right)\label{eq:20}
\end{align}
The specific mass flux radial momentum $\varPhi$ is defined as 
\[
\varPhi=r\rho_{0}v_{1}
\]
and its behaviour is given by the continuity equation \eqref{eq: 15}
\begin{equation}
\dot{\rho_{1}}+\frac{1}{r}\frac{\partial}{\partial r}\left(r\rho_{0}v_{1}\right)=\dot{\rho_{1}}+\frac{\varPhi^{\prime}}{r}=0\label{eq:21}
\end{equation}

\section{Solutions for homogeneous equations\label{sec:4 Solutions-for-homogeneous}}

\subsection{Time and space dependent separated equations\label{subsec:4.1 4.1 Time and space dependent separated equations}}

It seems hopeless to try to find an analytical solution to the third
order differential equation \eqref{eq:20} in $v_{1}$ and $\rho_{1}$.
However, a wave equation in $\rho_{1}$ with a mass term can be found
if one neglects the right side of \eqref{eq:20}: the gas is assumed
of low viscosity such as the viscous friction can be neglected in
front of the pressure gradient and of the central mass gravitational
gradient and secondly, the product of the radial derivatives of the
unperturbed specific mass $\rho_{0}$ and of the perturbed gravitational
potential $V_{1}$ is shown to be small (see Appendix C) and can be
neglected $\rho_{0}^{\prime}V_{1}^{\prime}\approx0$. Using notations
of\cite{Nowotny1979}, the equilibrium characteristics are written
with power law dependencies on the radial distance $r$. With the
dimensionless variable $R$, one defines
\begin{equation}
R=\frac{r}{r_{c}}\,\,\,;\,\,\rho_{0}=\rho_{c}R^{d}\,\,\,;\,\,c_{0}^{2}=c_{c}^{2}R^{s}\label{eq:22}
\end{equation}
where $r_{c}$ is a reference distance corresponding to the disc inner
radius, $\rho_{c}$ and $c_{c}$ are the nebula reference specific
mass and sound speed at the disc inner edge. The exponents $d$ and
$s$ depend on the nebula physical models and are addressed further.
The homogeneous equation \eqref{eq:20} becomes 
\begin{align}
 & \ddot{\rho_{1}}-\frac{R^{s}}{A^{2}}\left(\rho_{1}^{\prime\prime}+\left(2s+1-\frac{d+s}{\gamma}\right)\frac{\rho_{1}^{\prime}}{R}\right.\nonumber \\
 & \left.+\left(B^{2}R^{d+2-s}+s\left(s-\frac{d+s}{\gamma}\right)\right)\frac{\rho_{1}}{R^{2}}\right)=0\label{eq:23}
\end{align}
with, from now on, the prime sign $^{\prime}$ denoting $\partial\,/\partial R$
and where 
\[
A^{2}=\frac{r_{c}^{2}}{c_{c}^{2}}\,\,\,;\,\,B^{2}=\frac{4\pi G\rho_{c}r_{c}^{2}}{c_{c}^{2}}
\]
are constants. Posing 
\begin{align}
\rho_{1}\left(R,t\right) & =D\left(R\right)\Theta\left(t\right)\label{eq:24}\\
v_{1}\left(R,t\right) & =U\left(R\right)\varXi\left(t\right)\label{eq:24-1}\\
\varPhi\left(R,t\right) & =\Phi\left(R\right)\Psi\left(t\right)\label{eq:24-2}
\end{align}
and choosing $-\omega^{2}$ as separating constant ($\omega$ real),
for periodic perturbations that do not grow exponentially with time,
\eqref{eq:23} yields
\begin{equation}
\ddot{\Theta}\left(t\right)+\omega^{2}\Theta\left(t\right)=0\label{eq:25}
\end{equation}
\begin{align}
 & D^{\prime\prime}+\left(2s+1-\frac{d+s}{\gamma}\right)\frac{D^{\prime}}{R}+\nonumber \\
 & \left(B^{2}R^{d+2-s}+\omega^{2}A^{2}R^{2-s}+s\left(s-\frac{d+s}{\gamma}\right)\right)\frac{D}{R^{2}}=0\label{eq:26}
\end{align}
The perturbed continuity equation \eqref{eq: 15} yields, with $\kappa$
as a separating constant
\begin{equation}
\dot{\Theta}\left(t\right)-\kappa\Xi\left(t\right)=0\,\,\,;\,\,\Psi\left(t\right)=\Xi\left(t\right)\label{eq:27}
\end{equation}
\begin{equation}
U\left(R\right)=-\kappa\frac{r_{c}}{\rho_{c}}R^{-\left(d+1\right)}\int D\left(R\right)R\,dR\label{eq:28}
\end{equation}
\begin{equation}
\Phi\left(R\right)=r_{c}\rho_{c}R^{d+1}U\left(R\right)=-\kappa r_{c}^{2}\int D\left(R\right)R\,dR\label{eq:29}
\end{equation}
showing that $\Phi\left(R\right)$ is strongly dependent on the behaviour
of the radial perturbed velocity. 

The solutions of \eqref{eq:24} and \eqref{eq:26} for the time-dependent
part of $\rho_{1}$ and $v_{1}$ are
\begin{align}
 & \Theta\left(t\right)=C\sin\left(\omega t+\varphi\right)\label{eq:30}\\
 & \Xi\left(t\right)=\Psi\left(t\right)=\frac{C}{\kappa}\omega\cos\left(\omega t+\varphi\right)\label{eq:30-1}
\end{align}
with $C$ and $\varphi$ constants to be determined by initial conditions. 

The solutions \eqref{eq:30} show that the time dependent parts of
the gas perturbed specific mass $\Theta\left(t\right)$ and velocity
$\varXi\left(t\right)$ have the same frequency and the same initial
phase but they are out of phase by $\pi/2$, for $\kappa$ positive,
while the time dependent part of the specific mass flux radial momentum
$\Psi\left(t\right)$ is identical to the one of the gas perturbed
velocity $\varXi\left(t\right)$. The type of solution of equation
\eqref{eq:26} and hence the radial behaviour of $\rho_{1}$, $v_{1}$
and $\varPhi$ depend on the exponents $d$ and $s$ of the $\rho_{0}$
and $c_{0}$ radial distributions. Searching in the next sections
for analytical solutions of the equation \eqref{eq:26} for annular
structures to appear in the disc, we solve these equations \eqref{eq:26},
\eqref{eq:28} and \eqref{eq:29} for certain values of $d$ and $s$. 

Two boundary conditions are given: first, at the disc inner edge,
for $R=1$, the nebula perturbed specific mass must equal a parameter
$\rho_{c1}^{*}\left(t\right)$ independent of disc physical characteristics,
but that can depend on the time $t$, and second, for increasing $R$,
the nebula perturbed specific mass must decrease and vanish far away
from the central mass, for $R>>1$, for all time $t$.

The solutions for the perturbed azimuthal and vertical velocity components
are given in Appendix D.

\subsection{Solutions for d = 0 and s = 2 \label{subsec:4.2 4.2 Solutions for d =00003D 0 and s =00003D 2 }}

We consider first the unrealistic case of an uncompressible nebula
($d=0$) with a sound speed increasing linearly with the distance
($s=2$). This first case is purely theoretical, as for a nebula with
constant specific mass undergoing polytropic transformations, the
sound speed should be constant. The equation \eqref{eq:26} becomes
then a simple Euler type equation
\begin{align}
 & D^{\prime\prime}+\left(\frac{5\gamma-2}{\gamma}\right)\frac{D^{\prime}}{R}\label{eq:31}\\
 & +\left(B^{2}+\omega^{2}A^{2}+4\left(\frac{\gamma-1}{\gamma}\right)\right)\frac{D}{R^{2}}=0
\end{align}
Under the condition
\[
B^{2}+\omega^{2}A^{2}+4\left(\frac{\gamma-1}{\gamma}\right)>1
\]
yielding
\begin{equation}
\omega^{2}>\frac{c_{c}^{2}}{r_{c}^{2}}\left(\frac{4-3\gamma}{\gamma}\right)-4\pi G\rho_{c}\label{eq:32}
\end{equation}
and with the first boundary condition and posing
\[
y=\sqrt{B^{2}+\omega^{2}A^{2}+\frac{3\gamma-1}{\gamma}}
\]
the solution of \eqref{eq:31} reads
\begin{equation}
D=\frac{\rho_{c1}^{*}}{R}\cos\left(y\ln\left(R\right)\right)\label{eq:33}
\end{equation}
where $\ln$ is the Napier logarithm function. The radial terms of
the perturbed velocity and of the specific mass flux radial momentum
are found from \eqref{eq:28} and \eqref{eq:29}
\begin{align}
U & =-\kappa\frac{\rho_{c1}^{*}}{\rho_{c}}\frac{r_{c}}{y^{2}+1}R\cos\left(y\ln\left(R\right)-\arctan\left(y\right)\right)\label{eq:34}\\
\Phi & =-\kappa\rho_{c1}^{*}\frac{r_{c}^{2}}{y^{2}+1}R^{2}\cos\left(y\ln\left(R\right)-\arctan\left(y\right)\right)\label{eq:35}
\end{align}
The extrema (maxima and minima) of $D$ are found from 
\begin{equation}
D^{\prime}=-\frac{\rho_{c1}^{*}\sqrt{y^{2}+1}}{R^{2}}\cos\left(y\ln\left(R\right)-\arctan\left(y\right)\right)=0\label{eq:36}
\end{equation}
The zeros of $D$ \eqref{eq:33}, $U$\eqref{eq:34}, $\Phi$ \eqref{eq:35}
and $D^{\prime}$ \eqref{eq:36} are given by
\begin{align}
 & R=\alpha_{1}\left(\beta_{1}^{\backprime}\right)^{n}\label{eq:37}\\
 & \alpha_{1}=\exp\left(\frac{\pi/2+\varphi_{1}}{y}\right)\,\,\,;\,\,\beta_{1}^{\backprime}=\exp\left(\frac{\pi}{y}\right)\label{eq:37-1}
\end{align}
and $n$ non-negative integers, $\varphi_{1}=0$ for $D$ and $\varphi_{1}=\arctan\left(y\right)$
for $U$, $\Phi$ and $D^{\prime}$. The initial spatial phase between
$D$ and $U$ is $\arctan\left(y\right)=\pi/2$, provided that $y$
is large enough within the condition \eqref{eq:32}, while there is
no initial phase between $U$ (or $\Phi$) and $D^{\prime}$. The
distance ratio of two successive maxima of $D$, for $D^{\prime\prime}<0$,
is
\begin{equation}
\beta_{1}=\left(\beta_{1}^{\backprime}\right)^{2}=\exp\left(\frac{2\pi}{\frac{r_{c}}{c_{c}}\sqrt{\omega^{2}+4\pi G\rho_{c}+\frac{c_{c}^{2}}{r_{c}^{2}}\left(\frac{3\gamma-4}{\gamma}\right)}}\right)\label{eq:38}
\end{equation}
which, from \eqref{eq:32}, is a real constant depending on the nebula
characteristics $r_{c}$, $c_{c}$, $\rho_{c}$, $\gamma$ and on
the perturbations circular frequency $\omega$. Note that the condition
\eqref{eq:32} is equivalent to the dispersion relation in the classical
Jeans problem (see e.g., \cite{Tscharnuter1985}) with, for $\omega^{2}=0$,
critical wave number and wavelength
\begin{equation}
k_{crit}=\frac{\sqrt{4\pi G\rho_{c}}}{c_{c}}=\frac{\sqrt{\frac{4-3\gamma}{\gamma}}}{r_{c}}\,\,\,;\,\,\lambda_{crit}=2\pi r_{c}\sqrt{\frac{\gamma}{4-3\gamma}}\label{eq:39}
\end{equation}
The relation \eqref{eq:32} ensures that the perturbations do not
grow exponentially with time.

\subsection{Solutions for $s=2$ and $d<2(2\gamma-1)$, $d\protect\neq0$ \label{subsec: 4.3 Solutions for s=00003D2 and d<2(2=00005Cgamma-1), d=00005Cneq0 }}

In this second case, the sound speed increases linearly with the radial
distance and the specific mass depends on the radial distance, with
the conditions that $d$ must be non-null and smaller than $2(2\gamma-1)$.
The equation \eqref{eq:26} becomes
\begin{align}
 & D^{\prime\prime}+\left(\frac{5\gamma-\left(d+2\right)}{\gamma}\right)\frac{D^{\prime}}{R}\nonumber \\
 & +\left(B^{2}R^{d}+\omega^{2}A^{2}+2\left(\frac{2\gamma-\left(d+2\right)}{\gamma}\right)\right)\frac{D}{R^{2}}=0\label{eq:40}
\end{align}
which is a Bessel type equation, whose general solution reads
\begin{equation}
D=K_{1}R^{\left(\left(d+2\right)/2\gamma\right)-2}Z_{\nu}\left(z\right)\label{eq:41}
\end{equation}
where $Z_{\nu}\left(z\right)$ is the Bessel function of first kind
with $z$ the argument and $\nu$, from now on, the order
\begin{equation}
z=\frac{2}{d}B\,R^{d/2}\,\,\,;\,\,\nu=\frac{2}{d}\sqrt{\left(\frac{d+2}{2\gamma}\right)^{2}-\omega^{2}A^{2}}\label{eq:42}
\end{equation}
and $K_{1}$ is a constant determined by the first boundary condition
\[
K_{1}=\frac{\rho_{c1}^{*}}{Z_{\nu}\left(\frac{2}{d}B\right)}
\]
For circular frequencies $\omega$ such that
\begin{equation}
\omega>\frac{d+2}{2\gamma A}=\left(\frac{d+2}{2\gamma}\right)\frac{c_{c}}{r_{c}}\label{eq:43}
\end{equation}
the order $\nu$ is a pure imaginary, $\nu=jy$ with $j=\sqrt{-1}$
and, from now on,
\[
y=\frac{2}{d}\sqrt{\omega^{2}A^{2}-\left(\frac{d+2}{2\gamma}\right)^{2}}
\]
The function $Z_{\nu}\left(z\right)$ takes complex values and reads
generally \cite{Jahnke-Emde-Losch1966}
\begin{equation}
Z_{\nu}\left(z\right)=\left(\frac{z}{2}\right)^{\nu}\sum_{k=0}^{\infty}\frac{\left(-1\right)^{k}\left(\frac{z}{2}\right)^{2k}}{k!\,\Gamma\left(\nu+k+1\right)}\label{eq:44}
\end{equation}
where $\Gamma$ is the Legendre Gamma function. Writing
\begin{align}
 & \Gamma\left(k+1+jy\right)=h_{k}\exp\left(j\eta_{k}\right)\nonumber \\
 & h_{k}=k!\prod_{n=0}^{\infty}\frac{1}{\sqrt{\frac{y^{2}}{\left(k+1+n\right)^{2}}+1}}\label{eq:45}\\
 & \eta_{k}=y\Psi\left(k+1\right)\nonumber \\
 & +\sum_{n=0}^{\infty}\left(\frac{y}{\left(k+1+n\right)}-\arctan\left(\frac{y}{\left(k+1+n\right)}\right)\right)\nonumber 
\end{align}
where $\Psi$ is the digamma function, the Bessel function of imaginary
order reads
\begin{equation}
Z_{\nu}\left(z\right)=\sum_{k=0}^{\infty}C_{1k}\left(\frac{z}{2}\right)^{2k}\exp\left(j\left(y\ln\left(\frac{z}{2}\right)-\eta_{k}\right)\right)\label{eq:46}
\end{equation}
with
\[
C_{1k}=\exp\left(q\right)\frac{\left(-1\right)^{k}}{k!\,h_{k}}
\]
where $q=0$ if $d>0$ and $q=-\pi y$ if $d<0$ and where, from now
on, $z$ has to be replaced by its absolute value
\[
\left|z\right|=\frac{2}{\left|d\right|}B\,R^{d/2}
\]
Taking the real part of \eqref{eq:46}, the relation \eqref{eq:41}
reads
\begin{align}
D & =K_{1}R^{\left(\left(d+2\right)/2\gamma\right)-2}\sum_{k=0}^{\infty}\left[C_{1k}\left(\frac{z}{2}\right)^{2k}\right.\nonumber \\
 & \left.\cos\left(y\ln\left(\frac{z}{2}\right)-\eta_{k}\right)\right]\label{eq:47}
\end{align}
The second boundary condition, decreasing $D$ for increasing $R$,
restricts the exponent of $R$, giving the initial condition on $d$,
$d<2(2\gamma-1)$, $d\neq0$.

The radial terms of the perturbed velocity and of the specific mass
flux momentum become, from \eqref{eq:28} and \eqref{eq:29},
\begin{align}
U & =-\kappa K_{1}\frac{r_{c}}{\rho_{c}}R^{\left(\left(d+2\right)/2\gamma\right)-\left(d+1\right)}\sum_{k=0}^{\infty}\left[C_{2k}\left(\frac{z}{2}\right)^{2k}\right.\nonumber \\
 & \left.\sin\left(y\ln\left(\frac{z}{2}\right)-\eta_{k}+\tau_{k}\right)\right]\label{eq:48}\\
\Phi & =-\kappa K_{1}r_{c}^{2}R^{\left(d+2\right)/2\gamma}\sum_{k=0}^{\infty}\left[C_{2k}\left(\frac{z}{2}\right)^{2k}\right.\nonumber \\
 & \left.\sin\left(y\ln\left(\frac{z}{2}\right)-\eta_{k}+\tau_{k}\right)\right]\label{eq:49}
\end{align}
with
\begin{align*}
C_{2k} & =\frac{2C_{1k}}{\sqrt{\left(kd+\frac{d+2}{2\gamma}\right)^{2}+\left(\frac{yd}{2}\right)^{2}}}\\
\tau_{k} & =\arctan\left(\frac{2}{yd}\left(kd-\frac{d+2}{2\gamma}\right)\right)
\end{align*}
The extrema of $D$ are solutions of
\begin{align}
D^{\prime} & =-K_{1}R^{\left(\left(d+2\right)/2\gamma\right)-3}\sum_{k=0}^{\infty}\left[C_{3k}\left(\frac{z}{2}\right)^{2k}\right.\nonumber \\
 & \left.\sin\left(y\ln\left(\frac{z}{2}\right)-\eta_{k}+\mu_{k}\right)\right]=0\label{eq:50}
\end{align}
\begin{equation}
\tan\left(y\ln\left(\frac{z}{2}\right)\right)=\frac{\sum_{k=0}^{\infty}C_{3k}\left(\frac{z}{2}\right)^{2k}\sin\left(\eta_{k}-\mu_{k}\right)}{\sum_{k=0}^{\infty}C_{3k}\left(\frac{z}{2}\right)^{2k}\cos\left(\eta_{k}-\mu_{k}\right)}\label{eq:51}
\end{equation}
with
\begin{align*}
C_{3k} & =C_{1k}\sqrt{\left(2-kd-\frac{d+2}{2\gamma}\right)^{2}+\left(\frac{yd}{2}\right)^{2}}\\
\mu_{k} & =\arctan\left(\frac{2}{yd}\left(2-kd-\frac{d+2}{2\gamma}\right)\right)
\end{align*}
The zeros of $U$ and $\Phi$ are found like in \eqref{eq:51} with
$C_{3k}$ and $\mu_{k}$ replaced by $C_{2k}$ and $\tau_{k}$. It
seems that there are no simple analytical solutions to \eqref{eq:51}.
However, for small arguments $(z/2)<<1$ , i.e.,
\begin{equation}
\frac{4\pi G\rho_{c}r_{c}^{2}}{dc_{c}^{2}}R^{d}<<1\label{eq:52}
\end{equation}
one finds similar solutions for $D$ \eqref{eq:47}, $U$ \eqref{eq:48},
$\Phi$ \eqref{eq:49} and $D^{\prime}$ \eqref{eq:50}, in the form
\[
\tan\left(y\ln\left(\frac{z}{2}\right)\right)\approx\tan\left(\kappa\right)
\]
with $\kappa$ constant, as the first term for $k=0$ in the series
of \eqref{eq:51} predominates, yielding $\kappa=\eta_{0}$ for $D$,
$\kappa=\left(\eta_{0}-\tau_{0}\right)$ for $U$ and $\Phi$, and
$\kappa=\left(\eta_{0}-\mu_{0}\right)$ for $D^{\prime}$. 

The zeros of $D$ \eqref{eq:47}, $U$ \eqref{eq:48}, $\Phi$ \eqref{eq:49}
and $D^{\prime}$ \eqref{eq:50} are then given by 
\begin{align}
 & R=\alpha_{2}\left(\beta_{2}^{\backprime}\right)^{n}\label{eq:53}\\
 & \alpha_{2}=\left(\frac{\left|d\right|}{B}\right)^{2/d}\exp\left(\frac{2\left(\eta_{0}+\phi_{2}\right)}{dy}\right)\,\,\,;\,\,\beta_{2}^{\backprime}=\exp\left(\frac{2\pi}{dy}\right)\label{eq:53-1}
\end{align}
$n$ being non-negative integers and $\phi_{2}=\pi/2$ for $D$, $\phi_{2}=-\tau_{0}$
for $U$ and $\Phi$, and $\phi_{2}=-\mu_{0}$ for $D^{\prime}$.
Provided that $y$ is large enough within the condition \eqref{eq:43},
one has $\tau_{0}<<1$ and $\mu_{0}<<1$. The initial phase between
$D$ and $U$ is $\left(\pi/2\right)-\tau_{0}\approx\left(\pi/2\right)$,
while the initial phase between $U$ (or $\Phi$) and $D^{\prime}$
is $\left(\mu_{0}-\tau_{0}\right)\approx0$. The distance ratio of
two successive maxima of $D$ is
\begin{equation}
\beta_{2}=\left(\beta_{2}^{\backprime}\right)^{2}=\exp\left(\frac{2\pi}{\sqrt{\omega^{2}\frac{r_{c}^{2}}{c_{c}^{2}}-\frac{c_{c}^{2}}{r_{c}^{2}}\left(\frac{d+2}{2\gamma}\right)^{2}}}\right)\label{eq:54}
\end{equation}
which, from \eqref{eq:43}, is a real constant depending on nebula
reference characteristics and on $\omega$.

\subsection{Solutions for $d=s-2$ with $d>(2\gamma-1)/\left(1-\gamma\right)$,
$d\protect\neq0$ \label{subsec:4.4 Solutions for d=00003Ds-2 with d>(2=00005Cgamma-1)/=00005Cleft(1-=00005Cgamma=00005Cright), d=00005Cneq0 }}

The third case is more general and considers the two exponents linked
by the relation $d=s-2$ with the restrictions $d\neq0$ ($s\neq2$)
and $d>(2\gamma-1)/\left(1-\gamma\right)$. The equation \eqref{eq:26}
becomes 
\begin{align}
 & D^{\prime\prime}+\left(2d+5-\frac{2\left(d+1\right)}{\gamma}\right)\frac{D^{\prime}}{R}+\nonumber \\
 & \left(B^{2}+\frac{\omega^{2}A^{2}}{R^{d}}+\left(d+2\right)\left(d+2-\frac{2\left(d+1\right)}{\gamma}\right)\right)\frac{D}{R^{2}}=0\label{eq:55}
\end{align}
which is another Bessel type differential equation, whose solutions
are
\begin{equation}
D=K_{2}R^{\left(\left(d+1\right)/\gamma\right)-\left(d+2\right)}Z_{\nu}\left(z\right)\label{eq:56}
\end{equation}
where the argument $z$ and the order $\nu$ are now
\begin{equation}
z=\frac{2}{\left|d\right|}\omega AR^{\left|d\right|/2}\,\,\,;\,\,\nu=\frac{2}{\left|d\right|}\sqrt{\left(\frac{d+1}{\gamma}\right)^{2}-B^{2}}\label{eq:57}
\end{equation}
with $K_{2}$ a constant determined by the first boundary condition
\[
K_{2}=\frac{\rho_{c1}^{*}}{Z_{\nu}\left(\frac{2}{\left|d\right|}\omega A\right)}
\]
Under the condition
\[
B^{2}>\left(\frac{d+1}{\gamma}\right)^{2}
\]
yielding
\begin{equation}
\frac{4\pi G\rho_{c}r_{c}^{2}}{c_{c}^{2}}>\left(\frac{d+1}{\gamma}\right)^{2}\label{eq:58}
\end{equation}
the order $\nu$ is a pure imaginary, $\nu=jy$, with from now on
\[
y=\frac{2}{\left|d\right|}\sqrt{B^{2}-\left(\frac{d+1}{\gamma}\right)^{2}}
\]
Writing the Bessel functions of imaginary order as in \eqref{eq:46},
with $q=0$ in $C_{1k}$, the solution \eqref{eq:56} becomes
\begin{align}
D & =K_{2}R^{\left(\left(d+1\right)/\gamma\right)-\left(d+2\right)}\sum_{k=0}^{\infty}\left[C_{1k}\left(\frac{z}{2}\right)^{2k}\right.\nonumber \\
 & \left.\cos\left(y\ln\left(\frac{z}{2}\right)-\eta_{k}\right)\right]\label{eq:59}
\end{align}
The second boundary condition is fulfilled by the restriction on the
exponent of $R$ (with $\gamma>1$). 

The radial parts of the perturbed velocity and of the specific mass
flux momentum read, from \eqref{eq:28} and \eqref{eq:29},
\begin{align}
U & =-\kappa K_{2}\frac{r_{c}}{\rho_{c}}R^{\left(\left(d+1\right)/\gamma\right)-\left(2d+1\right)}\sum_{k=0}^{\infty}\left[C_{4k}\left(\frac{z}{2}\right)^{2k}\right.\nonumber \\
 & \left.\sin\left(y\ln\left(\frac{z}{2}\right)-\eta_{k}+\theta_{k}\right)\right]\label{eq:60}\\
\Phi & =-\kappa K_{2}r_{c}^{2}R^{\left(\left(d+1\right)/\gamma\right)-d}\sum_{k=0}^{\infty}\left[C_{4k}\left(\frac{z}{2}\right)^{2k}\right.\nonumber \\
 & \left.\sin\left(y\ln\left(\frac{z}{2}\right)-\eta_{k}+\theta_{k}\right)\right]\label{eq:61}
\end{align}
Like in the previous section, the extrema of $D$ are solutions of
\begin{align}
D^{\prime} & =-K_{2}R^{\left(\left(d+1\right)/\gamma\right)-\left(d+3\right)}\sum_{k=0}^{\infty}\left[C_{5k}\left(\frac{z}{2}\right)^{2k}\right.\nonumber \\
 & \left.\sin\left(y\ln\left(\frac{z}{2}\right)-\eta_{k}+\sigma_{k}\right)\right]=0\label{eq:62}
\end{align}
\begin{equation}
\tan\left(y\ln\left(\frac{z}{2}\right)\right)=\frac{\sum_{k=0}^{\infty}C_{5k}\left(\frac{z}{2}\right)^{2k}\sin\left(\eta_{k}-\sigma_{k}\right)}{\sum_{k=0}^{\infty}C_{5k}\left(\frac{z}{2}\right)^{2k}\cos\left(\eta_{k}-\sigma_{k}\right)}\label{eq:63}
\end{equation}
with
\begin{align*}
C_{5k} & =C_{1k}\sqrt{\left(d+2-\frac{d+1}{\gamma}-k\left|d\right|\right)^{2}+\left(\frac{yd}{2}\right)^{2}}\\
\sigma_{k} & =\arctan\left(\frac{2}{y\left|d\right|}\left(d+2-\frac{d+1}{\gamma}-k\left|d\right|\right)\right)
\end{align*}
The zeros of $U$ and $\Phi$ are found like in \eqref{eq:63} with
$C_{4k}$ and $\theta_{k}$ replacing $C_{5k}$ and $\sigma_{k}$. 

For small arguments $(z/2)<<1$, i.e.,
\begin{equation}
\omega<<\left|d\right|\frac{c_{c}}{r_{c}}R^{-\left|d\right|/2}\label{eq:64}
\end{equation}
one finds similar solutions for $D$ \eqref{eq:59}, $U$ \eqref{eq:60},
$\Phi$ \eqref{eq:61} and $D^{\prime}$ \eqref{eq:62} like in the
previous case, as \eqref{eq:63} is equal to a constant, $\tan\left(\kappa\right)$,
with $\kappa=\eta_{0}$ for $D$, $\kappa=\left(\eta_{0}-\theta_{0}\right)$
for $U$ and $\Phi$, and $\kappa=\left(\eta_{0}-\sigma_{0}\right)$
for$D^{\prime}$. 

The zeros of $D$ \eqref{eq:59}, $U$ \eqref{eq:60}, $\Phi$ \eqref{eq:61}
and $D^{\prime}$ read
\begin{align}
 & R=\alpha_{3}\left(\beta_{3}^{\backprime}\right)^{n}\label{eq:65}\\
 & \alpha_{3}=\left(\frac{\left|d\right|}{\omega A}\right)^{2/\left|d\right|}\exp\left(\frac{2\left(\eta_{0}+\phi_{3}\right)}{\left|d\right|y}\right)\,\,\,;\,\,\beta_{3}^{\backprime}=\exp\left(\frac{2\pi}{\left|d\right|y}\right)\label{eq:65-1}
\end{align}
with $n$ non-negative integers, $\phi_{3}=\pi/2$ for $D$, $\phi_{3}=-\theta_{0}$
for $U$ and $\Phi$, and $\phi_{3}=-\sigma_{0}$ for $D^{\prime}$.
Provided that $y$ is large enough within the condition \eqref{eq:58},
one has $\theta_{0}<<1$ and $\sigma_{0}<<1$ . The initial phase
between $D$ and $U$ (or $\Phi$) is $\pi/2-\theta_{0}\approx\pi/2$,
while the initial phase between $U$ (or $\Phi$) and $D^{\prime}$
is $\left(\sigma_{0}-\theta_{0}\right)\approx0$. 

The distance ratio of two successive maxima of $D$ is
\begin{equation}
\beta_{3}=\left(\beta_{3}^{\backprime}\right)^{2}=\exp\left(\frac{2\pi}{\sqrt{4\pi G\rho_{c}\frac{r_{c}^{2}}{c_{c}^{2}}-\left(\frac{d+1}{\gamma}\right)^{2}}}\right)\label{eq:66}
\end{equation}
which, from \eqref{eq:58}, is a real constant depending on the reference
characteristics but independent of $\omega$. The period of the small
perturbations must be larger than a minimum value
\begin{equation}
P_{m}=\frac{2\pi}{\left|d\right|}\frac{r_{c}}{c_{c}}\left(R_{max}\right)^{\left|d\right|/2}\label{eq:67}
\end{equation}
deduced from the condition \eqref{eq:64} applied to the whole range
of radial distances of a nebula ($R_{max}$ is the ratio of the disc
outer to inner radii).

\subsection{Standard model \label{subsec:4.5 Standard-model}}

We mention an interesting particular case, called the \textquotedbl standard
model\textquotedbl , of the general case $d=(s-2)$ above. One writes
the gravitational potential in the unperturbed disc as a power law
distribution in $R$ (= $r/r_{c}$) 
\begin{equation}
V_{0}=V_{c}R^{\upsilon}\label{eq:68}
\end{equation}
where $V_{c}$ is the gravitational potential of the central mass
$M^{*}$ (the gravitational potential of the disc is neglected as
$M_{d}<<M^{*}$) and $\upsilon$ is an exponent to be defined by physical
models. Replacing in the Poisson equation at equilibrium \eqref{eq:12}
with \eqref{eq:22} yields successively
\begin{align}
 & \frac{\upsilon^{2}V_{c}}{r_{c}^{2}}R^{\upsilon-2}=4\pi G\rho_{c}R^{d}\label{eq:69}\\
 & V_{c}=\frac{4\pi G\rho_{c}r_{c}^{2}}{\upsilon^{2}}=\frac{3GM_{c}}{\upsilon^{2}r_{c}}\label{eq:70}
\end{align}
for $d=\upsilon-2$ and with $M_{c}=\left(4\pi/3\right)r_{c}^{3}\rho_{c}$,
the mass of the homogeneous sphere of specific mass $\rho_{c}$ and
radius $r_{c}$ . 

We make the hypothesis for the \textquotedbl standard model\textquotedbl{}
that the reference distance $r_{c}$ of the disc inner edge can be
approximated by the central body unperturbed external radius $r_{c}^{*}$
\begin{equation}
r_{c}\approx r_{c}^{*}\label{eq:71}
\end{equation}
(superscript $^{*}$ denotes central body characteristics). Noting
the central body mean specific mass by $\rho^{*}$, identifying $V_{c}$
in \eqref{eq:70} with the gravitational potential of the central
mass $M^{*}$ yields
\begin{equation}
\rho_{c}=\frac{\upsilon^{2}}{3}\rho^{*}\label{eq:72}
\end{equation}
In the simplest case, the gravitational potential of a spherical body
is given by \eqref{eq:68}, with $\upsilon=-1$. The condition \eqref{eq:70}
yields then $d=-3$ and, from \eqref{eq:72}, the nebula reference
specific mass $\rho_{c}$ is one third of the mean specific mass of
the central body. 

On the other hand, within the perfect gas approximation, the sound
speed distribution \eqref{eq:22} follows the gas temperature radial
distribution in the disc, which can be represented by a power law
relation of exponent $\zeta$
\begin{equation}
c_{c}^{2}R^{s}=\frac{\gamma\Re}{\mu}T_{c}R^{\zeta}\label{eq:73}
\end{equation}
with $\Re$ the perfect gas constant, $\mu$ the gas molecular mass
and $T_{c}$ a reference temperature at the disc inner edge, that
can be approximated for example by the central body effective temperature.
The radial behaviour of the temperature in a nebula is model dependent.
Considering only the central body luminosity as the dominant source
of energy heating the nebula (the gas viscosity is neglected), the
temperature gradient is adiabatic with $\zeta=-1$ for an optically
thick nebula \cite{Pollack_etal1977}, yielding $s=-1$. 

We define then the \textquotedbl standard model\textquotedbl{} of
a disc as the case with $\upsilon=-1$, $d=-3$ and $s=-1$, and it
can be solved with these values by the general case $d=(s-2)$ above.
The distance ratio of maxima of the gas perturbed specific mass distribution
writes then, from \eqref{eq:66},
\begin{equation}
\beta_{st.mod.}=\exp\left(\frac{2\pi c_{c}}{\sqrt{\frac{GM^{*}}{r_{c}}-\left(\frac{2c_{c}}{\gamma}\right)^{2}}}\right)\label{eq:74}
\end{equation}
The condition \eqref{eq:58} ensures that this ratio is a real constant.

This simple \textquotedbl standard model\textquotedbl{} can be useful
as a first approximation model, provided that the disc mass $M_{d}$
calculated with the value \eqref{eq:72} of $\rho_{c}$ fulfills the
initial condition $M_{d}<<M^{*}$. Let's note also that in the above
approximation, the value $\rho_{c1}^{*}\left(t\right)$ that the nebula
perturbed specific mass has to match at the disc inner edge (first
boundary condition) can be approximated by the perturbed specific
mass of the central body at its outer edge, for $r=r_{c}^{*}=r_{c}$
or $R=1$, at the epoch $t$. (Strictly speaking, one should consider
the central body external perturbed radius $r_{c1}^{*}=r_{c}^{*}+\xi\left(r_{c}^{*},t\right)$,
where $\xi\left(r_{c}^{*},t\right)$ is the radial displacement of
the central body outer edge at $r=r_{c}^{*}$ due to small perturbations
at the epoch $t$, yielding $R_{c1}=r_{c1}^{*}/r_{c}^{*}=1+\xi/r_{c}^{*}$;
but if the displacements are small in front of the central body unperturbed
radius ($\xi<<r_{c}^{*}$), one has $R_{c1}\approx R_{c}=1$).

\section{Formation of annular structures \label{sec:5 5 Formation of annular structures }}

For all the cases considered, the spatial part of the perturbed specific
mass $D$ has a sign opposite to the signs of its radial derivative
$D^{\prime}$, of the radial perturbed velocity $U$ and of the specific
mass flux radial momentum $\Phi$. The functions $U$ and $\Phi$
have an initial phase difference of approximately $\pi/2$ with respect
to the function $D$. The zeros of $U$ correspond to the extrema
of $D$ and vice-versa. For increasing $R$, $U$ and $\Phi$ are
positive (respectively negative) between successive minima and maxima
(respectively successive maxima and minima) of $D$, as shown in Figure
1 of \cite{Pletser2022}. This configuration yields radial outward
flows of gas between successive minima and maxima of $D$ and radial
inward flows of gas between successive maxima and minima. The extrema
amplitudes of $D$ and $D^{\prime}$ decrease for increasing $R$,
while the extrema amplitudes of $U$ and $\Phi$ increase for increasing
$R$, although less for $\Phi$ than for $U$ in the case $d=(s-2)$.
The nebular gas, flowing outward (respectively inward) with a positive
(respectively negative) radial velocity $U$, may accumulate in annular
rings centered on circular orbits with radii corresponding to the
distances of the maxima of the gas perturbed specific mass, depleting
the zones of minima of perturbed specific mass. 

In a rotating nebula containing \textquotedbl dust\textquotedbl ,
the solid particles experience an inward drift due to the gas drag
caused by the difference of the gas circular velocity and the Keplerian
orbital velocity, the former being less than the latter \cite{Weidenschilling1977}.
Smaller particles are more affected by the gas drag than larger ones.
Particles on eccentric orbits encounter gas of variable density, causing
a circularization of their orbit. If a radial velocity is superimposed
onto the gas circular velocity, solid particles experience an additional
radial drag causing the orbit of smaller particles to decay more (respectively
less) rapidly in the case of inward (respectively outward) gas flow,
larger particles being less affected. The nebular \textquotedbl dust\textquotedbl{}
is dragged along with the gas, causing the orbits eccentricity of
particles to change, favouring collision and accretion (see e.g.,
\cite{Brahic1977}). This process would eventually result in an accumulation
of solid particles dragged along with the gas, near zones of maxima
of gas perturbed specific mass. A more detailed analysis of the dynamical
gas/particle interactions would confirm this, but is outside the scope
of this paper.

\section{Conclusions\label{sec:6 Conclusions}}

It was shown that, when under small radial periodic perturbations
and disregarding non-radial perturbations, thin slowly rotating low
mass gaseous discs, described by a simple two-dimensional axisymmetric
model, evolve such as the perturbed part of the gas specific mass
displays exponentially spaced maxima, two successive maxima being
separated by a minimum. The gas flows from locations of specific mass
minima inward to the preceding maximum or outward to the next maximum,
as the gas radial velocity is negative (inward flow) or positive (outward
flow). This mechanism would eventually form gaseous annular structures. 

Furthermore, the distance ratio of two successive maxima is found
to be a constant depending on disc characteristics (and on the perturbations
frequency for the first two cases). The nature and origin of the perturbations
are not discussed here. However, one can make the hypothesis that
the origin of the perturbations may lie within the central mass or
at the interface disc/central mass, due to periodic radial motions. 

Lower limit on orders of magnitudes of time scales can be deduced
for the case $d=(s-2)$ from the condition \eqref{eq:67} on the period
of the perturbations. For the \textquotedbl standard model\textquotedbl ,
minimum periods depend on dimensions of the central mass and are in
the order of several $10^{3}$ years for protostellar discs similar
to what the protoplanetary disc around the proto-Sun may have been
and in the order of several $10^{-1}$ year for giant planets proto-satellite
discs. In a second paper, we explore analytical solutions of the perturbed
specific mass wave-like propagation by considering two other general
cases.
\begin{acknowledgements}
We wish to thank Prof. O. Godart, Catholic University of Louvain,
Louvain-la-Neuve, Belgium, for early discussions on the subject of
this paper, Prof. P. Paquet, Catholic University of Louvain, for guidance
during this research work, and Dr D. Pletser and Dr C. Byrne for their
hospitality in Oxford during final redaction.
\end{acknowledgements}

\section*{Appendix A\label{sec:Appendix-A}}

The specific mass flux due to the radial periodic perturbation can
be divided in two parts. The radial part is due to the perturbed radial
velocity $v_{1}$ and reads $\rho_{0}v_{1}$ while the azimuthal part
has two components, the first one due to the azimuthal velocity at
equilibrium $v_{0}$ multiplied by the perturbed specific mass $\rho_{1}$
and the second one due to the perturbed azimuthal velocity $u_{1}$
multiplied by the specific mass $\rho_{0}$ at equilibrium, that is
$\rho_{1}v_{0}+\rho_{0}u_{1}$. We show here that the contribution
of the second term $\rho_{0}u_{1}$ to the azimuthal specific mass
flux is in fact much smaller than the first one $\rho_{1}v_{0}$ and
can be neglected.

We show first that $u_{1}$ is much smaller than $v_{0}$. Under the
hypothesis of purely axisymmetric radial perturbations, all perturbed
variables are function of the radius $r$ and time $t$. So, the perturbed
azimuthal velocity $u_{1}$ depends only on $r$ and $t$ and not
on the azimuthal angle $\theta$. Therefore, $u_{1}$does not appear
in the continuity equation. However, we still can find a relation
between $u_{1}$ and $v_{0}$.

The perturbed azimuthal velocity $u_{1}$ is related to the radial
perturbed velocity $v_{1}$ by the Coriolis effect. With $\Omega$
the norm of the nebula rotation angular velocity vector $\boldsymbol{\Omega}$
pointing upward, the Coriolis acceleration vector has a norm $-2\Omega v_{1}$
and is in the azimuthal direction of $v_{0}$ if $v_{1}$ is directed
radially inward and in the opposite azimuthal direction of $v_{0}$
if $v_{1}$ is directed radially outward. As the perturbations are
purely radial and periodic, let $\omega$ be the angular frequency
and the perturbed radial position $r_{1}=\varepsilon\sin\left(\omega t\right)$,
with the amplitude $\varepsilon$ much smaller than the radial position
$\varepsilon<<r$, then the perturbed radial velocity reads $v_{1}=\varepsilon\omega\cos\left(\omega t\right)\leq\varepsilon\omega$,
yielding a periodically changing Coriolis acceleration $a_{c1}=-2\Omega\varepsilon\omega\cos\left(\omega t\right)$. 

The perturbed azimuthal velocity $u_{1}$ is then in the order of
$u_{1}\approx\int a_{c1}dt=-2\Omega\varepsilon\sin\left(\omega t\right)\leq2\Omega\varepsilon$.
The azimuthal velocity at equilibrium $v_{0}$ is in the order of,
or less than, $\Omega r$ (see \ref{sec:Appendix-B}). Then the ratio
\begin{equation}
\frac{u_{1}}{v_{0}}\leq\frac{2\Omega\varepsilon}{\Omega r}=\frac{2\varepsilon}{r}<<1\label{eq:A1}
\end{equation}
and the azimuthal velocity during perturbation is $v_{0}+u_{1}=v_{0}\left(1+\frac{u_{1}}{v_{0}}\right)\approx v_{0}$.

Furthermore, as the perturbations are periodic, the second term of
the azimuthal specific mass flux is $\rho_{0}u_{1}=-2\rho_{0}\Omega\varepsilon\sin\left(\omega t\right)$
and is varying relatively fast as $\omega>>\Omega$, i.e., its azimuthal
direction changes sense relatively quickly between opposite and along
the unperturbed velocity $v_{0}$. Its average contribution $\left\langle \rho_{0}u_{1}\right\rangle $
over a period $T=\frac{2\pi}{\omega}$ is therefore small in front
of the larger contribution of the first term $\rho_{1}v_{0}$ and
can be neglected. That is $\rho_{1}v_{0}+2\left\langle \rho_{0}u_{1}\right\rangle =\rho_{1}v_{0}\left(1+2\frac{\left\langle \rho_{0}u_{1}\right\rangle }{\rho_{1}v_{0}}\right)\approx\rho_{1}v_{0}$.

\section*{Appendix B\label{sec:Appendix-B}}

Solving for the gas velocity $v_{0}$ at equilibrium within the hypothesis
that the kinematic viscosity is negligible ($\nu=0$), the azimuthal
component of the Navier-Stokes equation \eqref{eq:11} yields
\begin{equation}
v_{0}^{\prime\prime}+\frac{v_{0}^{\prime}}{r}-\frac{v_{0}}{r^{2}}=f_{v}(r)\label{eq:B1}
\end{equation}
with the notation $"^{\,\prime}"=\partial\,/\partial r$ and where
$f_{v}(r)$ is an unspecified function of $r$, giving in general
\begin{equation}
v_{0}=C_{1}\frac{1}{r}+C_{2}r+F_{v}(r)\label{eq:B2}
\end{equation}
with 
\[
F_{v}(r)=\frac{2}{r^{2}}\int\left(\intop f_{v}(r)dr\right)rdr
\]
and $C_{1}$ and $C_{2}$ constants determined by boundary conditions.
If the viscosity $\nu$ is non null, then obviously $f_{v}(r)$ and
$F_{v}(r)$ have to be nil in \eqref{eq:B1} and \eqref{eq:B2}. For
$r\rightarrow\infty$, the gas circular velocity has to stay within
finite values, yielding theoretically $C_{2}=0$. Another expression
of the gas circular velocity $v_{0}$ at equilibrium is found from
the radial component of the Navier-Stokes equation \eqref{eq:10}.
Using \eqref{eq:17} and \eqref{eq:22}, it yields
\begin{equation}
v_{0}=\sqrt{\frac{GM^{*}}{r}+\frac{c_{c}^{2}}{r_{c}^{s}}\left(\frac{s+d}{\gamma}\right)r^{s}}\label{eq:B3}
\end{equation}
where $s$ and $d$ are usually negative. For the gas circular velocity
$v_{0}$ to be real, the Keplerian velocity has to be greater than
the velocity induced by the gas gradient pressure, which is usually
the case in real nebulae \cite{Weidenschilling1977}. The relations
\eqref{eq:B2} and \eqref{eq:B3} are complementary in describing
the radial profile of the circular gas velocity at equilibrium. For
$\nu\neq0$ ($F_{v}(r)=0$) and noting generally $v_{0}\left(r\right)\sim r^{p}$,
the value $p=+1$ ($C_{1}=0$) gives the rotation velocity of a solid,
and approximately of a fluid with high viscosity, at a constant angular
speed. A value $p=-1$ ($C_{2}=0$) describes the rotation of a perfect
gas, and approximately of a fluid with low viscosity. The value $p=-1/2$
corresponds to a Keplerian rotation. A value $p=s/2$ describes the
rotation of a gas dominated by thermal pressure. The gas circular
velocity profile in a real nebula is at least a combination of the
three first cases, as observed in the galaxies' rotation curves \cite{Shapley1972,Bowers_=000026_Deeming1984}:
highly viscous fluid near the central mass ($v_{0}\left(r\right)$
$\approx$ linear relation), lesser viscous fluid further from the
centre ($v_{0}\left(r\right)$ $\approx$ inverse linear relation)
and, after a transition region, approximate Keplerian rotation in
the external regions ($v_{0}\left(r\right)$ $\approx$ inverse root
square relation).

\section*{Appendix C\label{sec:Appendix-C}}

One can neglect $\rho_{0}^{\prime}V_{1}^{\prime}$ in \eqref{eq:20}
if small displacements occur due to small radial perturbations. Assuming
that a fluid element is displaced from vectorial positions $x$ to
$x+\xi\left(x\right)$, where $\xi\left(x\right)$ is a small displacement,
vectorial function of $x$, the perturbed specific mass at $x$ reads

\begin{equation}
\rho_{1}\left(x\right)=-\nabla.\left(\rho_{0}\xi\right)\label{eq:C1}
\end{equation}
where the specific mass in the divergence operator is replaced by
the unperturbed specific mass as it is multiplied by the small quantity
$\xi$ \cite{Chandrasekhar1969,Binney_Tremaine1987}. Assuming that
$\rho_{1}$ and $\xi$ depend only on $r$ in a cylindrical polar
referential, $\xi=\left(\xi\left(r\right),0,0\right)$), the relation
\eqref{eq:C1} reads
\begin{equation}
\rho_{1}\left(r\right)=\frac{-1}{r}\frac{\partial\left(r\rho_{0}\xi\right)}{\partial r}\label{eq:C2}
\end{equation}
and from \eqref{eq:9} and \eqref{eq:6}, with the notation $"^{\,\prime}"=\partial\,/\partial r$,
\begin{equation}
V_{1}^{\prime}=\frac{4\pi G}{r}\int\rho_{1}r\,dr=-4\pi G\rho_{0}\xi\label{eq:C3}
\end{equation}
The product $\rho_{0}^{\prime}V_{1}^{\prime}$ in \eqref{eq:20} reads
then, with \eqref{eq:22},
\begin{equation}
\rho_{0}^{\prime}V_{1}^{\prime}=-4\pi Gd\rho_{c}^{2}R^{2d-1}\left(\frac{\xi}{r_{c}}\right)\label{eq:C4}
\end{equation}
showing that it can be neglected if the small displacement $\xi$
is small enough in comparison with the central mass radius $r_{c}$.

\section*{Appendix D\label{sec:Appendix-D}}

The perturbed azimuthal velocity is found from \eqref{eq:14} and
reads now

\begin{equation}
\dot{u_{1}}+\frac{v_{1}}{r_{c}}\left(v_{0}^{\prime}+\frac{v_{0}}{R}\right)=0\label{eq:D1}
\end{equation}

yielding successively, with \eqref{eq:24} and \eqref{eq:30}, $\kappa_{1}$
as (negative) separating constant, and using $u_{1}=0$ at $t=0$
as initial condition, 
\begin{align}
u_{1} & =\frac{\kappa_{1}}{r_{c}}\left(v_{0}^{\prime}+\frac{v_{0}}{R}\right)\int v_{1}dt\nonumber \\
 & =\frac{\kappa_{1}}{r_{c}}\left(v_{0}^{\prime}+\frac{v_{0}}{R}\right)U\left(R\right)\int\varXi\left(t\right)dt\nonumber \\
 & =\frac{\kappa_{1}C}{\kappa r_{c}}\omega\left(v_{0}^{\prime}+\frac{v_{0}}{R}\right)U\left(R\right)\int\cos\left(\omega t\right)dt\nonumber \\
 & =\frac{\kappa_{1}C}{\kappa r_{c}}\left(v_{0}^{\prime}+\frac{v_{0}}{R}\right)U\left(R\right)\sin\left(\omega t\right)\label{eq:D2}
\end{align}

showing that the perturbed azimuthal velocity $u_{1}$ is periodic
by nature.

The perturbed vertical velocity \eqref{eq:14-1} reads now $\dot{w_{1}}=0$,
yielding with $w_{1}=0$ at $t=0$ as initial condition, that the
perturbed vertical velocity is nil at all time, $w_{1}=0$.

\end{document}